\documentclass[preprint]{aastex}
\usepackage{latexsym}
\usepackage{natbib}

\newcommand{\sbeam}[5]{{$#1\mbox{$''\mskip-7.6mu.\,$}#2$} $\times$ {$#3\mbox{$''\mskip-7.6mu.\,$}#4$}; P.A.=$#5^\circ$}

\newcommand{\msec}[2]{$#1\mbox{$''\mskip-7.6mu.\,$}#2$}
\newcommand{\Msun}{$M_\odot$}

\shorttitle{Expansion parallax of the planetary nebula IC 418}
\shortauthors{Guzman et al.}

\begin{document}

\title{Expansion parallax of the planetary nebula IC 418}

\author{Lizette Guzm\'an\altaffilmark{1}, Laurent Loinard, Yolanda G\'omez}
\affil{Centro de Radioastronom\'ia y Astrof\'isica, Universidad Nacional Aut\'onoma de M\'exico,\\  58089 Morelia, Michoac\'an, M\'exico}
\and
\author{Christophe Morisset}
\affil{Instituto de Astronom\'ia, Universidad Nacional Aut\'onoma de M\'exico,\\ 04510 M\'exico, D.F., M\'exico}

\altaffiltext{1}{Also at: Jodrell Bank Centre for Astrophysics, 
University of Manchester,
Manchester, M13 9PL, UK}

\begin{abstract}

In this paper, we present radio continuum observations of the 
planetary nebula IC 418 obtained at two epochs separated by
more than 20 years. These data allow us to show that the angular 
expansion rate of the ionization front in IC 418 is 5.8 $\pm$ 1.5 
mas~yr$^{-1}$. If the expansion velocity of the ionization front is 
equal to the expansion velocity of the gas along the line of sight 
as measured by optical spectroscopy, then the distance to IC 418 
must be 1.1 $\pm$ 0.3 kpc. Recent theoretical predictions 
appropriate for the case of IC 418, however, suggest that the 
ionization front may be expanding about 20\% faster than 
the material. Under this assumption, the distance to IC 418 
would increase to 1.3 $\pm$ 0.4 kpc.

\end{abstract}


\section{Introduction}

The large uncertainties affecting distance estimates to planetary nebulae 
(PNe) continue to represent an important obstacle to our understanding
of these objects. While the total number of PNe in the Milky Way has been 
estimated to be somewhere between 5,000 and 25,000 (Peimbert 1990, 
Zijlstra \& Pottasch 1991, Acker et al.\ 1992), less than 50 have distances 
that have been measured individually with reasonable accuracy. About
a third of these individual estimates are direct trigonometric parallax
measurements (Harris et al.\ 2007), while the rest is based on a disparate
set of indirect methods (e.g.\ spectroscopic parallax of the central star or 
of a resolved companion, cluster membership, reddening, or angular 
expansion). In addition, statistical methods, calibrated using PNe with 
individually measured distances, have been developed. A frequently 
used such statistical technique (called the Shklovsky method) is based 
on the assumption that the mass of ionized gas is the same for all PNe. 
This rather crude assumption has been refined by Daub (1982) and 
Cahn et al.\  (1992) who distinguish between density-bounded PNe (for 
which the Shklovsky constant mass hypothesis is assumed to hold) and 
radiation-bounded PNe (where the mass of ionized gas is taken to depend 
on the surface brightness of the nebula). For a recent review of various
methods of distance determinations to PNe, see Phillips (2002).

A potentially reliable direct geometric method to estimate the distance of
an individual PN is the so-called expansion parallax technique. In this method, the angular
expansion of the source on the plane of the sky is measured, and
compared to the expansion along the line of sight determined from
optical spectroscopy. Traditionally,  the angular expansion has been 
obtained from multi-epoch radio interferometric data gathered over 
a period of several years (Masson 1986), but measurements at optical 
wavelengths with the Hubble Space Telescope have also been used 
(Reed et al.\ 1999; Palen et al.\ 2002). This technique has been applied successfully 
to several PNe (Masson 1986, 1989; G\'omez et al.\ 1993; Hajian et al.\ 1993, 1995; 
Kawamura \& Masson 1996; Hajian \& Terzian 1996; Christianto \& Seaquist 1998; 
Guzman et al.\ 2006),
and can provide distances accurate to about 20\%. This is comparable
to the accuracy obtained from existing  trigonometric parallax
measurements (Harris et al.\ 2007), and significantly better than 
the accuracy of statistical methods (e.g.\ Stanghellini et al.\ 2008). 
In this paper, we will apply the expansion parallax technique to 
the well studied PN IC 418. 

\begin{deluxetable}{lcccc}
\tabletypesize{\scriptsize}
\tablecaption{List of observations.\label{tbl-1}}
\startdata
\hline
Project & Epoch & Phase Calibrator & Bootstrapped Flux Density(Jy) & Restored Beam\\
\tableline
AP116 & 1986 Jun 28 (1986.49) & 0605-085 & 2.347$\pm$0.008 & \sbeam{3}{80}{3}{08}{0}  \\
AL716 & 2007 Nov 06 (2007.85) & 0516-160 & 0.425$\pm$0.005 & \sbeam{3}{80}{3}{08}{0}  \\
\enddata
\end{deluxetable}

The morphology of IC 418 (G215.2-24.2, PK 215-24.1, The Spirograph Nebula) is fairly simple: both 
at optical and radio wavelengths, it has an elliptical ring shape, with
a major axis of 14$''$ and a minor axis of 10$''$. It is surrounded by a 
low-level ionized halo, which is itself enshrouded in a neutral envelope 
with an angular size of about 2$'$ (Taylor \& Pottasch 1987, Taylor et al.\
1989). Widely discrepant estimates of the distance to IC 418 have been 
obtained using different statistical methods. To our knowledge, the 
shortest distance ever proposed is 360 pc (Acker 1978)
whereas the largest one is 5.74 kpc (Phillips \& Pottasch 1984). 
The Shklovsky method mentioned earlier gives a value of 1.9 kpc 
(Cahn \& Kaler 1971). In recent years, the most popular value appears 
to have been 1 kpc (e.g.\ Meixner et al.\ 1996, Pottasch et al.\ 2004), 
although the reason for this is not entirely clear. 

\begin{figure*}[h!!]
\centering
\includegraphics[scale=0.80, angle=0]{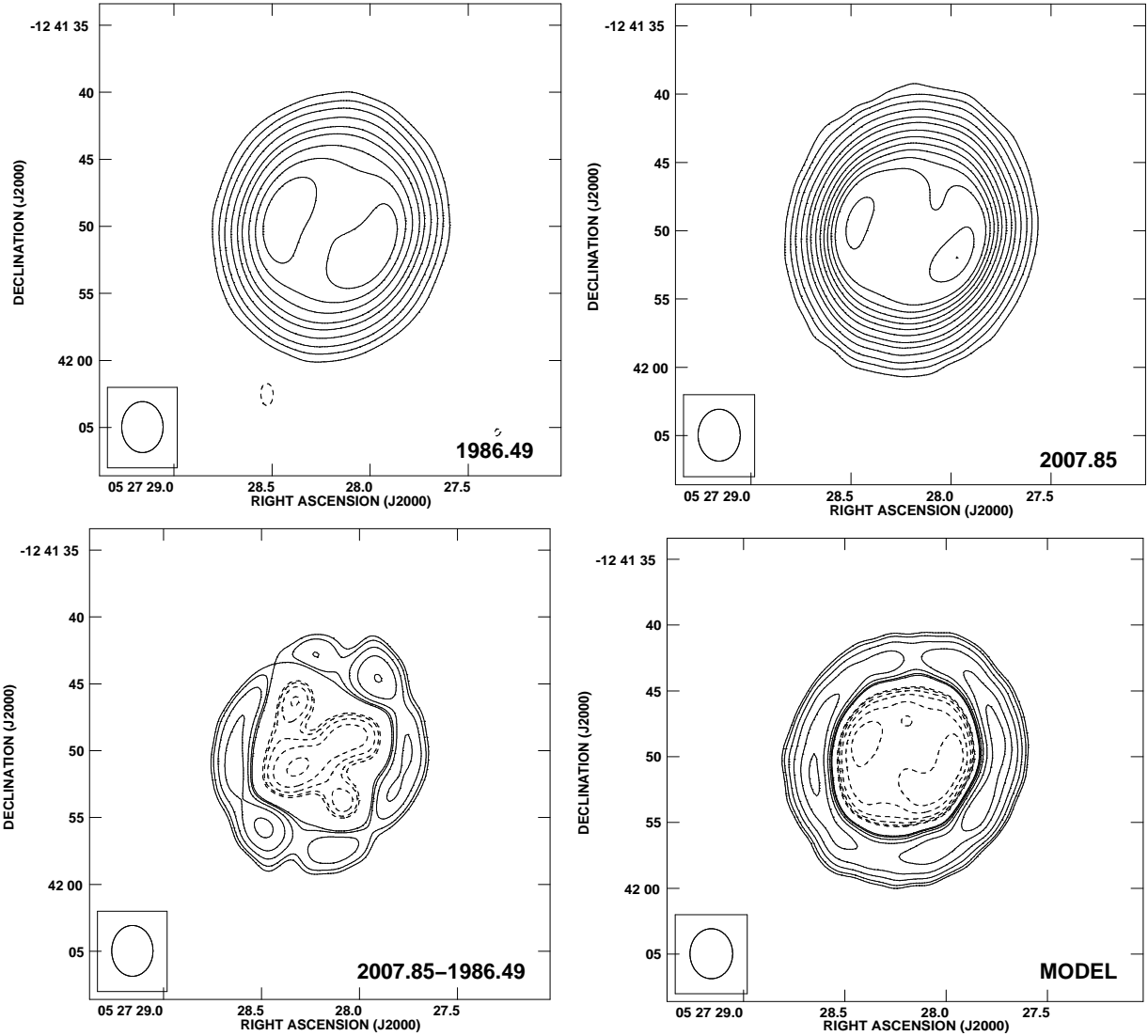}
 \caption{\footnotesize{Top: contour images of the 6 cm continuum
emission from IC 418 for 1986.49 (left) and 2007.85 (right).
The contours are  -5, 5, 10, 30, 50, 60, 80, 100, 150, 200, 250, 300 
times 350 $\mu$Jy, the average rms noise of the images. 
Bottom: contour images of the 6 cm difference image
(left) and of the ``model'' (right) obtained as described in the text.
The contours are -20, -15, -10, -7, -5, -4, 4, 5, 7, 
10, 15, and 20 times 460 $\mu$Jy, the rms noise of the difference image.
The restoring beam ($3\rlap.{''}80 \times 3\rlap.{''}08$
with a position angle of $0^\circ$) is shown in the bottom left corner
of each image.}}
\label{fig1}
\end{figure*}

\section{Observations}

The data (Tab.\ 1) were collected with the Very Large Array 
(VLA) of the NRAO\footnote{The National Radio Astronomy 
Observatory is operated by Associated Universities, Inc. under a 
cooperative agreement with the Nacional Science Foundation.} 
at 6 cm (5 GHz) in its second most extended (B) configuration
on 1986, June 28 (1986.49) and 2007, November 06 (2007.85). 
This provides a time separation of  21.36 yr. The data were edited 
and calibrated using the Astronomical Image Processing System 
(AIPS) following standard procedures  (see Table 1). The longest 
baselines were tapered down to increase the sensitivity, and 
self-calibration (in phase only) was applied. The resulting images
are shown in Fig.\ 1.

\section{Results and discussion}

The difference image between the two epochs (new minus old) was 
produced following Guzman et al.\ (2006). It is shown in the bottom left panel 
of Fig.\ 1, and exhibits the typical concentric negative/positive pattern 
expected when expansion is present. It is possible to estimate the 
expansion rate of the nebula by comparing the difference image of 
the data with a model. That model can be built from the data of either 
epochs, but the best results are obtained using the dataset with the
highest quality (corresponding to the second epoch in the present
case). To generate the model difference, 
we image the data twice. The first imaging was made using a
fixed pixel size  of \msec{0}{1}, whereas  the second was made 
using pixels of $(1 + \epsilon) \times $ \msec{0}{1}, with 
$\epsilon$ $\geq$ 0, but $\epsilon$ $\ll$ 1. We then formed the 
pixel-to-pixel subtraction between those two images. The physical 
size of the source, of course, does not depend on the chosen
size of the pixel. But since the pixels are larger in the second case,
the source occupies a smaller number of them. Therefore, the 
pixel-to-pixel subtraction is equivalent to subtracting from the real image 
of the source, a self-similarly shrunk (i.e.\ a de-magnified) version of itself. 
To identify the best 
model, we repeated the procedure described above for a set of
values of $\epsilon$. For each value, we compared the model
difference with the real difference between the second and the
first epoch. The best model clearly corresponds to the situation where
the two differences are as similar to each other as possible. To obtain
a quantitative measure of when that happens, we subtracted the two 
difference images, and calculated the r.m.s.\ of the resulting image.
A plot of that r.m.s.\ as a function of $\epsilon$ (Fig.\ 2) shows that 
the minimum of the r.m.s.\  corresponds to a well-defined value of 
$\epsilon$. That value was calculated using a fit to the data points 
with a quadratic form (blue curve in Fig.\  3). This yields a value 
of $\epsilon$ = $0.018 \pm 0.005$.

\begin{figure*}[h!!]
\centering
\includegraphics[scale=0.3, angle=0]{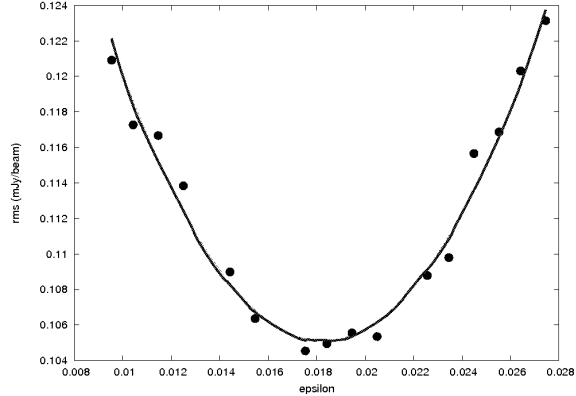}
 \caption{\footnotesize{Residual r.m.s.\ value as a function of $\epsilon$. 
 The blue line shows the best quadratic fit.}}
\label{fig3}
\end{figure*}

The angular expansion rate can be determined from the value of 
$\epsilon$ using:

$$ \dot{\theta} = {{\theta~ \epsilon} \over {\Delta t}}.$$

\noindent The radius of maximum emission, $\theta$, is estimated from
the image at the second epoch to be $6\rlap.{''}700 \pm 0\rlap.{''}006$, 
so $\dot{\theta} = 5.8 \pm 1.5$ mas~yr$^{-1}$. Since the radio emission 
considered here traces the ionized region of IC 418, $\dot{\theta}$ 
corresponds to the expansion rate of the ionization front.

To deduce the distance from the angular expansion rate calculated 
above, one must know the physical velocity $v_{exp}$ at which the 
ionization front is expanding (e.g.\ Guzman et al.\ 2006):

$$\bigg[\frac {D} {pc}\bigg] = 211 \bigg[\frac {v_{exp}} {km~s^{-1}}\bigg] \bigg[ \frac {\dot{\theta}} {mas~yr^{-1}} \bigg]^{-1}.$$

\noindent 
Traditionally, $v_{exp}$ has been estimated using high spectral 
resolution observations of some emission lines and assuming a 
relation between the shape of the line profiles and the movement 
of the emitting gas. In the present case, we use the high resolution 
spectra of H$\beta$, [N II] and [O III] lines published by Gesicki et al.\  (1996). 
The H$\beta$ line is broad (FWHM = 18 km s$^{-1}$, dominated 
by the thermal contribution), while the [O III] line is narrow and 
centered on the systemic velocity. The [N II] line, on the other hand, 
is double-peaked with each peak at $\pm$ 20 km s$^{-1}$ 
around the systemic velocity. Using a detailed 3D photoionization model 
of the nebula, Morisset and Georgiev (in prep.) reproduced these three 
profiles using a $V$ $\propto$ $R^4$ expansion law. According to this model, the
expansion velocity near the outer edge of the nebula (where we detect 
expansion at radio wavelengths) is 30 km s$^{-1}$. This result is in good 
agreement with the radiation-hydrodynamic models of Sch{\" o}nberner
et al.\ (2005) and Villaver et al.\ (2002).  With this value of $v_{exp}$ = 
30 km~s$^{-1}$, we obtain a distance to IC 418 of 1.1 $\pm$ 0.3 kpc.

There is a potential systematic effect discussed in detail by Mellema 
(2004) that ought to be taken into account. While the angular expansion 
measured using the VLA data corresponds to the progression of the 
ionization front (and is, therefore, a pattern velocity), the Doppler width 
deduced from spectral lines traces the expansion velocity of the 
material itself. For conditions appropriate for PNe, Mellema (2004) 
showed that  the ionization front tends to expand somewhat faster 
than the material. Using their Fig.\ 5, we estimate that the 
expansion velocity of the ionization front in IC 418 is 1.2 $\pm$ 0.1 faster
than the expansion velocity of the gas. Taking this effect into account
leads to a distance to IC 418 of 1.3 $\pm$ 0.4 kpc.

These two estimates of the distance are well within the broad range
(0.36 to 5.74 kpc) of values obtained from statistical methods (see
Sect.\ 1). They are also in very good agreement with the value of 
1 kpc that has often been used recently, and with the value foung
by Morisset \& Georgiev (in prep.): $d$ = 1.26 $\pm$ 0.2 kpc\footnote{
Morisset and Georgiev perform a detailed modeling of both the star 
and the nebula of IC418. They reproduce more than 140 nebular lines, 
as well as the HST images of the nebula, and the electron temperature 
and density diagnostics. All the main stellar emission and absorption 
lines are also reproduced. In this model, there is a degeneracy between the distance, the 
absolute magnitude, the size of the nebula, and the presence of clumps. They use 
evolutionary tracks to resolve the degeneracy and conclude that the filling factor 
is one. This leads to the distance of 1.26 $\pm$ 0.2 kpc mentioned here.}.

Two distance independent parameters can be derived from our
observations. First, the dynamical age of the nebula can be calculated from
the angular size and angular expansion rate: $\tau_{dyn} = \theta/\dot{\theta}$. 
We obtain $\tau_{dyn}$ $\sim$~1,200 years. Second, the emission
measure of the ionized region can be obtained from the observed parameters
of the radio emission (we used the image at the second epoch, which is
of better quality; the total 6 cm flux is 1.44$\pm$0.01 Jy and the average 
deconvolved angular size of the emission is 
13${\rlap .}^{\prime\prime}$4$\pm$0${\rlap .}^{\prime\prime}$01). This
yields an emission measure EM = (5.05 $\pm$ 0.01) $\times$ 10$^{-6}$
cm$^{-6}$ pc. Finally, the electron density and the mass of ionized
gas can be calculated (also from the radio flux). We get
$n_e$ = (6.2 $\pm$ 1.7) $({d \over 1.1 kpc})^{-0.5}$ $\times$  10$^3$ cm$^{-3}$, 
and $M_i$ = (8.7 $\pm$ 2.4) $({d \over 1.1 kpc})^{2.5}$ $\times$  10$^{-2}$ \Msun.
This is in reasonable agreement with the values found by Morisset \& Georgiev 
(in prep.) from their detailed modeling ($n_e$ = 9 $\times$ 10$^3$ cm$^{-3}$ and
$M_i$ = 6 $\times$  10$^{-2}$ \Msun).

\section{Conclusions}

In this paper, we presented observations of the 6 cm radio continuum
emission from the well-studied planetary nebula IC 418 obtained at two 
epochs separated by more than 20 yr. These data allowed us to detect 
the angular expansion of the nebula, and to estimate its distance. Depending
on the assumption made on the relative velocity of the matter and of
the ionization front, we obtain a distance of 1.1 $\pm$ 0.3 kpc, or 1.3 $\pm$ 
0.4 kpc.

\acknowledgments

LG, YG and LL aknowledge the support of DGAPA, UNAM and CONACYT (M\'exico).
This research has made use of the SIMBAD database, operated at CDS, Strasbourg, France.

\end{document}